\documentclass[aps,prl,twocolumn,notitlepage,superscriptaddress,longbibliography]{revtex4-1}  
\setcitestyle{square}
\usepackage{graphicx}  
\usepackage{float}
\usepackage{dcolumn}   
\usepackage{bm}        
\usepackage{amssymb}   
\usepackage{amsmath}
\usepackage{isomath}
\usepackage[usenames, dvipsnames]{color}
\usepackage{soul}
\usepackage{physics}
\usepackage[normalem]{ulem}
\usepackage{xpatch}
\usepackage{siunitx}	
\usepackage{xcolor}
\usepackage{hyperref}
\hypersetup{
     colorlinks   = false,
     citecolor    = blue
}
\begin{document}

\widetext
\title{Gradient-index granular crystals: From boomerang motion to \\ asymmetric transmission of waves}

\author{Eunho Kim}
\affiliation{Aeronautics and Astronautics, University of Washington, Seattle, WA, USA, 98195-2400}
\affiliation{Division of Mechanical System Engineering, Jeonbuk National University, 567 Baekje-daero, Deokjin-gu, Jeonju-si, Jeollabuk-do, Republic of Korea, 54896}
\affiliation{Automotive Hi-Technology Research Center \& LANL-CBNU Engineering Institute-Korea, Jeonbuk National University, 567 Baekje-daero, Deokjin-gu, Jeonju-si, Jeollabuk-do, Republic of Korea, 54896}

\author{Rajesh Chaunsali}
\affiliation{Aeronautics and Astronautics, University of Washington, Seattle, WA, USA, 98195-2400}

\author{Jinkyu Yang}
\affiliation{Aeronautics and Astronautics, University of Washington, Seattle, WA, USA, 98195-2400}
\email{jkyang@aa.washington.edu}

\date{\today}

\begin{abstract}
We present a gradient-index crystal that offers extreme tunability in terms of manipulating the propagation of elastic waves. For small-amplitude excitations, we achieve control over wave transmission depth into the crystal. We numerically and experimentally demonstrate a boomerang-like motion of wave packet injected into the crystal. For large-amplitude excitations on the same crystal, we invoke nonlinear effects. We numerically and experimentally demonstrate asymmetric wave transmission from two opposite ends of the crystal. Such tunable systems can thus inspire a novel class of designed materials to control linear and nonlinear elastic wave propagation in multi-scales.
\end{abstract}
\maketitle

\textit{Introduction}.--- 
The advent of phononic crystals and metamaterials in recent years have shown excellent possibilities to manipulate elastic waves in materials~\citep{Maldovan2013, Kadic2013, Hussein2014}. 
Several ingenious designs have been proposed to build exotic devices, e.g., diode~\citep{Li2004, Liang2009, Boechler2011}, cloak~\citep{Farhat2009}, negative refraction metamaterial~\citep{Zhang2004, Zhu2014},  energy harvester~\citep{Yang2004, Carrara2013}, impact absorber~\citep{Daraio2006}, flow stabilizer~\citep{Hussein2015} and topological lattice~\citep{Wang2015, Ma2019}. 
The key idea is to use one or more ingredients among structural periodicity~\citep{Brillouin1953}, local resonances~\citep{Liu2000}, nonlinear effects~\citep{Chong2017}, etc., to achieve nontrivial dynamical responses. 
The underlying physics in these demonstrations could also open new ways to control mechanical vibrations at nanoscale by optomechanical~\citep{Eichenfield2009} and nanophononic metamaterials~\citep{Davis2014}.
Therefore, the need of exploring advanced material architectures that offer rich wave physics is ever growing.

In this context, granular crystals~\citep{Nesterenko2001, Sen2008} -- systematic arrangement of granular particles -- offer a unique advantage. These architectures mimic atomic lattice dynamics in the sense that the grains can be regarded as atoms that interact via a nonlinear interaction potential stemming from the nature of the contact.
These crystals are highly tunable, and a plethora of wave physics can be demonstrated in the same system~\citep{Porter2015}. 
Control over wave propagation in this setting shows many technological advances, ranging from impact and blast protection~\citep{Kim2015} to micro-scale granular beds~\citep{Hiraiwa2016}.

In this Letter, we present a \textit{gradient-index} granular crystal that offers even further tunability in terms of manipulating both linear and nonlinear elastic waves. These granular particles are of a cylindrical shape, where simply by tuning the contact angles between them, a gradient in stiffness can be achieved~\citep{Khatri2012}. Gradient-index materials have been extensively studied in optics and acoustics for various purposes, such as rainbow trapping~\citep{Tsakmakidis2007}, opening wide bandgaps~\citep{Kushwaha1998}, waveguides~\citep{Kurt2007, He2008}, lens~\citep{Smith2005, Torrent2007, Horsley2014, Jin2019}, beamwidth compressor~\cite{Lin2009a}, wave concentration~\citep{Romero-Garcia2013}, and absorbers~\citep{Climente2012, Liang2014}. Gradient-index systems are unique as the gradual variation in material/structural properties enables control over wave speed and wave directions at the same time minimizing wave scattering. 

Using the gradient-index granular crystal, here, we numerically and experimentally demonstrate capability of wave control in two fronts. For small-amplitude waves, the system follows linear dynamics, and therefore, we demonstrate frequency-dependent wave penetration into the system. This includes a boomerang-like motion of injected wave packet that returns back to the point of excitation without propagating through the whole crystal. This is similar to a mirage effect.
For large-amplitude waves, we invoke nonlinear effects \citep{Narisetti2010, Ganesh2013, Abedinnasab2013}, and we show that the system offers asymmetric wave transmission in two opposite directions. This leads to one-way energy transport as a result of the interaction of nonlinearity and spatial asymmetry~\citep{Yang2007, Liang2009, Boechler2011, Devaux2015, Wu2018, Moore2018}. Remarkably, all these characteristics can be tuned simply by changing the stacking angles and controlling the wave amplitude in the system.

 \begin{figure}[t]
\centering
\includegraphics[width=3.0in]{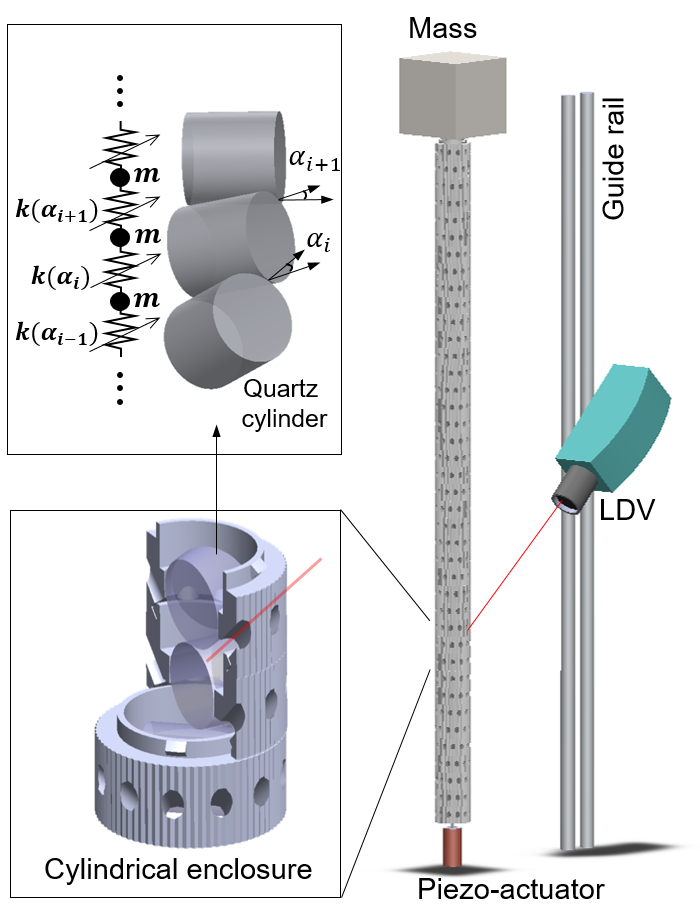}
\caption{(Color online) Experiment setup to investigate wave dynamics in gradient-index granular crystal. 
}
\label{fig1}
\end{figure}

 \begin{figure*}[t]
\centering
\includegraphics[width=7in]{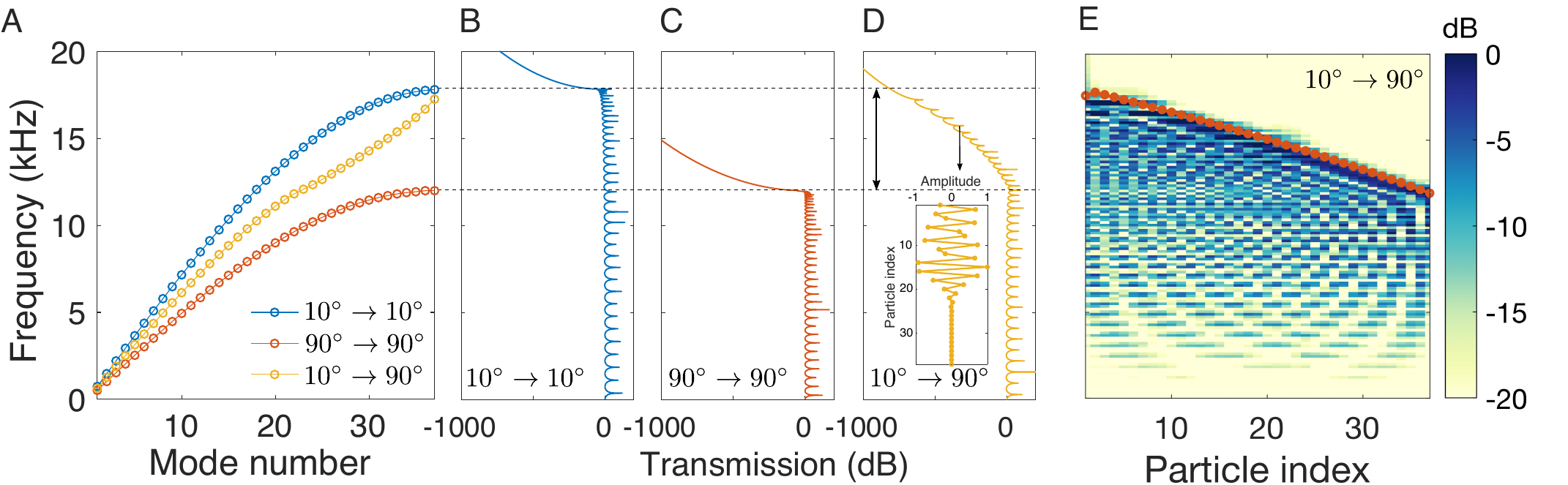}
\caption{(Color online) Modal frequencies and transmission of the gradient-index granular chain in comparison to homogeneous chains. (A) Modal frequencies of a gradient-index chain with contact angle $10^{\circ} \rightarrow 90^{\circ}$ along with the homogeneous chains with only $10^{\circ}$ and $90^{\circ}$. (B)-(C) State-space-based wave transmission for the homogeneous chain with contact angle $10^{\circ}$ and $90^{\circ}$. (D) The same for the gradient-index chain. (E) Frequency spectrum vs. space obtained from full numerical simulations on the gradient-index chain under an impact excitation. The highlighted line in red represents analytically obtained \textit{local} cutoff frequency.}
\label{fig2}
\end{figure*} 

\textit{Experimental and numerical setup}.---Our system is composed of 37 cylinders (with the length and the diameter equal to 18 mm) stacked vertically and pre-compressed by a free weight on top as shown in Fig.~\ref{fig1}. We vertically align the cylinders using 3D-printed cylindrical enclosures (lower inset of Fig.~\ref{fig1}). Each enclosure has one cylinder inside, and deliberate clearances are provided to restrict their rattling in rotation and to minimize any friction in the translational direction. The enclosures are assembled in series and can be rotated independently to dial in contact angles between neighboring cylindrical particles inside. The cylinders interact as per the Hertz contact law~\citep{Johnson1985}, and therefore, linear (nonlinear) wave dynamics can be studied at small (large) dynamic excitations in comparison to the static pre-compressive force ($F_0=29.4$ N).  
We vary the contact angles ranging from 10$^{\circ}$ to 90$^{\circ}$ along the chain such that the contact stiffness varies linearly along the chain. 10$^{\circ}$ represents a stiff side, whereas 90$^{\circ}$ is a soft side. A piezoelectric actuator (Piezomechanik PSt 500/10/25 VS18) is placed at the bottom of the chain in contact with the first particle. The actuator excites the chain using a Gaussian wave packet with a specific central frequency. A function generator (Agilent 33220A) sends the input to the actuator via an amplifier (Piezomechanik LE 150/100 EBW). We measure velocity of each particle by a laser Doppler vibrometer (Polytec OFV-534) at $45^{\circ}$ through the delicately-designed holes in the enclosures. 
The point-by-point measurements of the particles are synchronized to reconstruct the wave field along the chain.

To investigate wave dynamics, we first numerically model the system by employing the discrete element method (upper inset of Fig.~\ref{fig1}). Each fused-quartz cylinder (Young's modulus $E=72$ GPa, Poisson's ratio $\nu =0.17$, and density $\rho=2200$ kg/m$^3$) is considered as a point mass having only one degree of freedom in the vertical direction. The interaction of $i$th and $(i+1)$th cylinders -- making a contact angle $\alpha_i$ -- is modeled as the following force-displacement law: $F=\beta(\alpha_i)(\delta_{i} + u_i -u_{i+1})^{3/2}$. Here $\beta(\alpha_i)$ is the contact stiffness coefficient, $u_i$ denotes the dynamic displacement of $i$th cylinder, and $\delta_{i}$ is the pre-compression due to the static force given to the system (see Supplementary Material \citep{Suppl} for the full expression of $\beta(\alpha_i)$ and equations of motion).
We neglect the gravitational force, because it is much smaller compared to $F_0$. We explore the linear wave dynamics of the system by studying modal response of the system. To this end, for small dynamical excitations, we can linearize our contact model such that contact stiffness $k_{\text{lin}}(\alpha_i)=(3/2)\beta(\alpha_i)^{2/3}F_0^{1/3}$. 

\textit{Linear dynamics}.---In Fig.~\ref{fig2}A, we show modal frequencies of lossless gradient-index chain ($10^{\circ} \rightarrow 90^{\circ}$) in comparison to \textit{homogeneous} chains ($10^{\circ} \rightarrow 10^{\circ}$ and $90^{\circ} \rightarrow 90^{\circ}$), i.e., uniform contact angle (thus stiffness $k_{\text{lin}}$) along the length. 
We observe that the eigen frequencies of the $10^{\circ} \rightarrow 10^{\circ}$ chain span upto a cutoff frequency about 17.78 kHz [$=(1/\pi)  \sqrt{k_{lin}(10^{\circ})/m}$, where $m$ represents the mass of cylinders and $k_{lin}(10^{\circ})$ denotes linearized stiffness for $10^{\circ}$ contact]. Similarly, for the $90^{\circ} \rightarrow 90^{\circ}$ chain, we observe eigen frequencies cover the spectrum upto 11.97 kHz [$=(1/\pi) \sqrt{k_{lin}(90^{\circ})/m}$]. 
For the gradient-index chain, however, we observe eigen frequencies extend to about 17.78 kHz (i.e., the cutoff frequency for $10^{\circ} \rightarrow 10^{\circ}$ chain), but the curve has a portion that is concave upward starting from about 11.97 kHz (the cutoff frequency for $90^{\circ} \rightarrow 90^{\circ}$ chain). These modes are referred to as ``gradons'' in the previous literature~\citep{Xiao2006}.

 \begin{figure*}[t]
\centering
\includegraphics[width=6.6in]{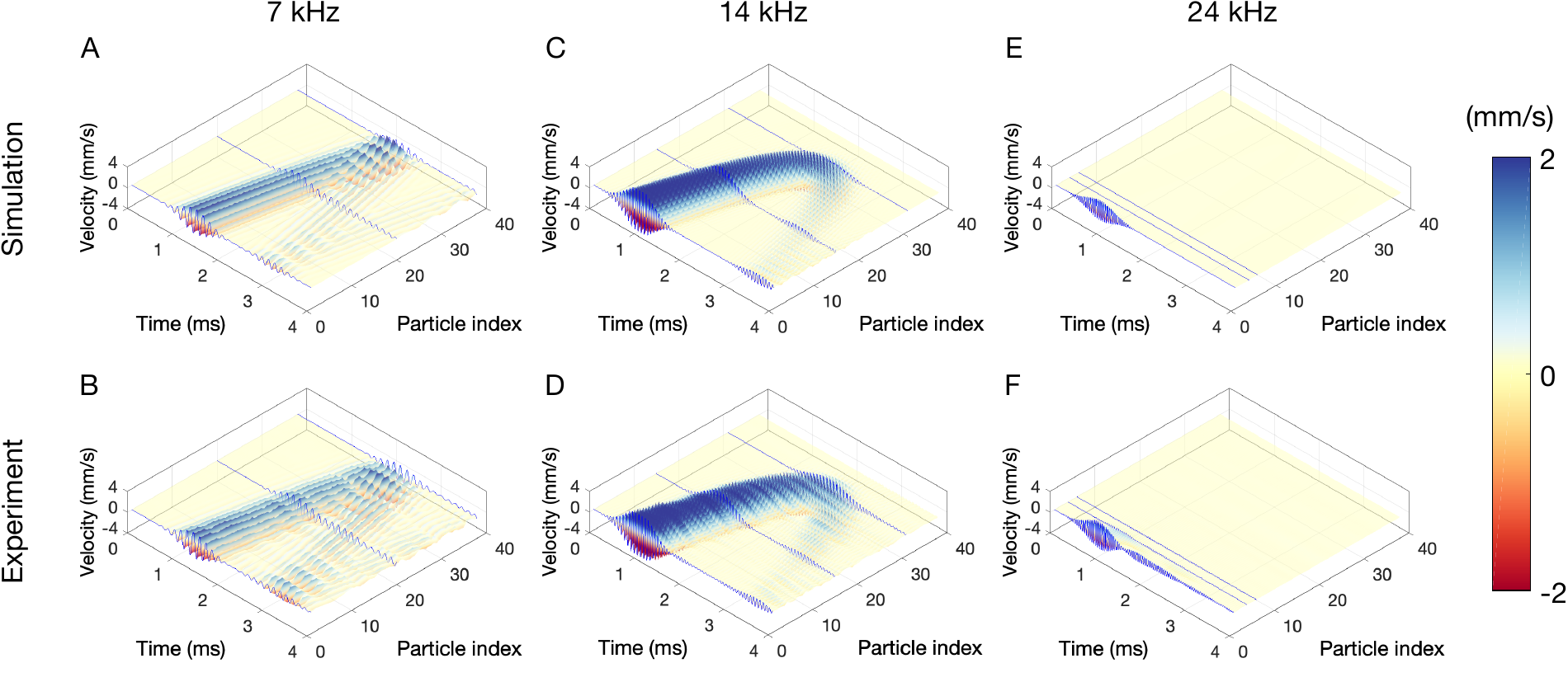}
\caption{(Color online) Linear wave dynamics in the gradient-index chain ($10^{\circ} \rightarrow 90^{\circ}$) under a Gaussian-modulated wave excitation. (A)-(B) Numerically- and experimentally-obtained spatiotemporal velocity maps for the excitation centered at 7 kHz.
3D line plots (in blue) are superimposed to highlight velocity time history at certain locations along the chain. 
(C)-(D) The same at 14 kHz with the boomerang-like wave propagation. (E)-(F) The same at 24 kHz.}
\label{fig3}
\end{figure*}

To investigate further, we plot the wave transmission as a function of frequency for all the aforementioned configurations in Figs.~\ref{fig2}B-D. For this, we use state-space approach to calculate the ratio of output force (felt by the upper mass) to input force (by the lower actuator) \citep{Boechler2011a}. It is evident that $10^{\circ} \rightarrow 10^{\circ}$ and $90^{\circ} \rightarrow 90^{\circ}$ homogeneous chains have pass bands from 0 kHz to their respective cutoff frequencies, whereas the gradient chain shows a pass band with decreasing transmission in the frequency range marked by the double-sided arrow in Fig.~\ref{fig2}D, which corresponds to the region with the concave upward trend in Fig.~\ref{fig2}A. We show in the inset a mode shape for a frequency in this region. Since its modal amplitude dominate the chain only partially, we can explain why the wave transmission decreases in this region. 

We verify this argument further by performing full numerical simulation with a small-amplitude impulse excitation given to the chain (Runge-Kutta solver with 0.01 m/s of initial velocity to the first particle). We then perform the fast Fourier transformation (FFT) on the velocity time history of each particle to plot frequency spectrum along the length of the chain as shown in Fig.~\ref{fig2}E. We observe that the wave transmission is only partial along the chain in the frequency range mentioned above. As the input frequency increases in this region, the transmission is more limited to the front end of the chain. Therefore, we can interpret linear dynamics in this gradient-index chain as if the system has spatially-varying ``local'' cutoff frequency. Analytical expression of such a local cutoff frequency can be mathematically expressed as  $f_{c,i}=(1/\pi) \sqrt{k_{lin}(\alpha_i)/m}$), which shows an excellent fit with the numerical results shown Fig.~\ref{fig2}E. We thus conclude that the our gradient-index chain would have three regions of wave transmission. From 0 kHz to 11.97 kHz, there is a pass band; from 11.97 kHz to 17.78 kHz, there is a \textit{quasi} stop band, i.e., wave transmission upto a fraction of the chain; and for frequencies above 17.78 kHz, there is a stop band.

With the understanding of the three different regions of wave transmission in our gradient-index chain, we now send Gaussian-modulated waveforms centered at frequencies residing in these three regions. We numerically and experimentally show how the wave packet propagates along the chain when sent from the stiffer side ($10^{\circ}$).
In Figs.~\ref{fig3}A and B, we show spatiotemporal evolution of a wave packet at 7 kHz obtained numerically and experimentally. As the frequency falls in the region of full transmission, we clearly observe that the wave packet is transmitted to the other end of chain. A significant decay in amplitude, however, is due to the damping in the experiments, which is modeled in simulations as well (Supplementary Material \citep{Suppl}). In Figs.~\ref{fig3}C and D, we show spatiotemporal evolution of a wave packet at 14 kHz, which lies in the partial wave transmission region. Evidently, the wave packet slows down as it propagates along the chain. It stops at a spatial location and then turns back to the front of the chain. This is analogous to boomerang motion, which we could successfully capture in our experiments. This boomerang motion typically involves wave amplification near the turning location (Supplementary Material \citep{Suppl}). Lastly, the wave sent at 24 kHz in the stop band does not propagate along the chain at all and is confined to the left end (Figs.~\ref{fig3}E and F). In this way, we have demonstrated that our gradient-index system offers a great control over the penetration depth of the wave packet as a function of its frequency.

\begin{figure}[t]
\centering
\includegraphics[width=3.7in]{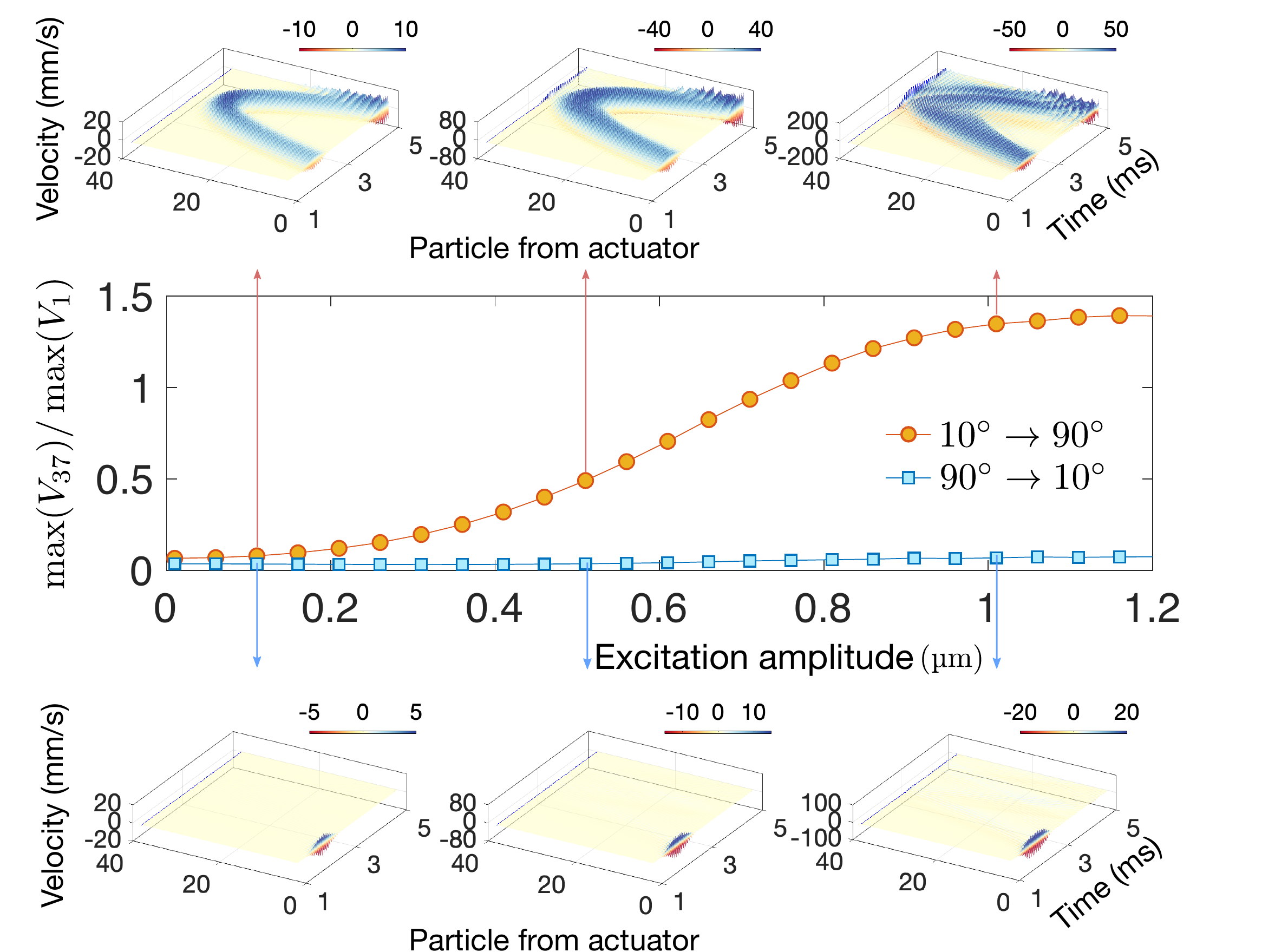}
\caption{(Color online) Nonlinear wave dynamics in the gradient-index granular crystal. We observe asymmetry in wave transmission with increasing excitation amplitude from \SI{0.11}{\micro\metre} (left panel), \SI{0.51}{\micro\metre} (middle), to \SI{1.01}{\micro\metre} (right), at 13.5 kHz for two configurations: forward ($10^{\circ} \rightarrow 90^{\circ}$, top panel) and reverse ($90^{\circ} \rightarrow 10^{\circ}$, bottom). }
\label{fig4}
\end{figure}

\textit{Nonlinear dynamics}.---We now investigate wave dynamics for larger amplitudes by invoking nonlinear effects.
In particular, we consider the frequency regime that offers partial wave transmission, the uniqueness of this gradient-index chain, and then increase wave amplitude to assess transmission characteristics of the system. 
We send a Gaussian-modulated pulse centered at 13.5 kHz from the two opposite ends and numerically monitor wave transmission as shown in Fig.~\ref{fig4}.
We quantify wave transmission as the ratio of the maximum velocity of the last particle to that of the first particle. Viscous damping is ignored here. 
For small-amplitude  excitations, the forward configuration ($10^{\circ} \rightarrow 90^{\circ}$) shows boomerang wave motion and returns back without reaching the other end as predicted earlier. However, upon increasing the wave amplitude, we see significant rise in wave transmission through the chain due to wave leakage as seen in the upper panels of Fig.~\ref{fig4}. In contrast, for the reverse configuration ($90^{\circ} \rightarrow 10^{\circ}$), the wave does not penetrate the bulk of the chain and remains localized near the excitation point as seen in the bottom panels of Fig.~\ref{fig4}. Upon increasing the wave amplitude, the localization still persists, and there is not a significant rise in the wave transmission.

This amplitude-dependent asymmetric wave transmission can be understood as the interplay between nonlinearity and spatial gradient in the system. Looking back at the eigenmode (``gradon'') plotted in the inset in Fig.~\ref{fig2}D, when we excite the system from the stiffer side ($10^{\circ}$), the presence of larger modal amplitude contributes to invoking nonlinear effects (such as frequency shifts) easily with an increased excitation amplitude. However, when we excite the system from the soft side ($90^{\circ}$), nonlinear effects become substantially suppressed, similar to the mechanism observed in thermal systems~\citep{Yang2007}. By further investigating this phenomenon in the frequency spectra for both small and large excitation amplitudes, we observe that the enhancement of the wave transmission in the forward configuration is due to the \textit{gradual} frequency softening and spatial extension of nonlinear mode (Supplementary Material \citep{Suppl}). We note that this mechanism is different from those relying on harmonic generation \citep{Liang2009}, bifurcation \citep{Boechler2011}, or self-demodulation \citep{Devaux2015}.

\begin{figure}[t]
\centering
\includegraphics[width=3.5in]{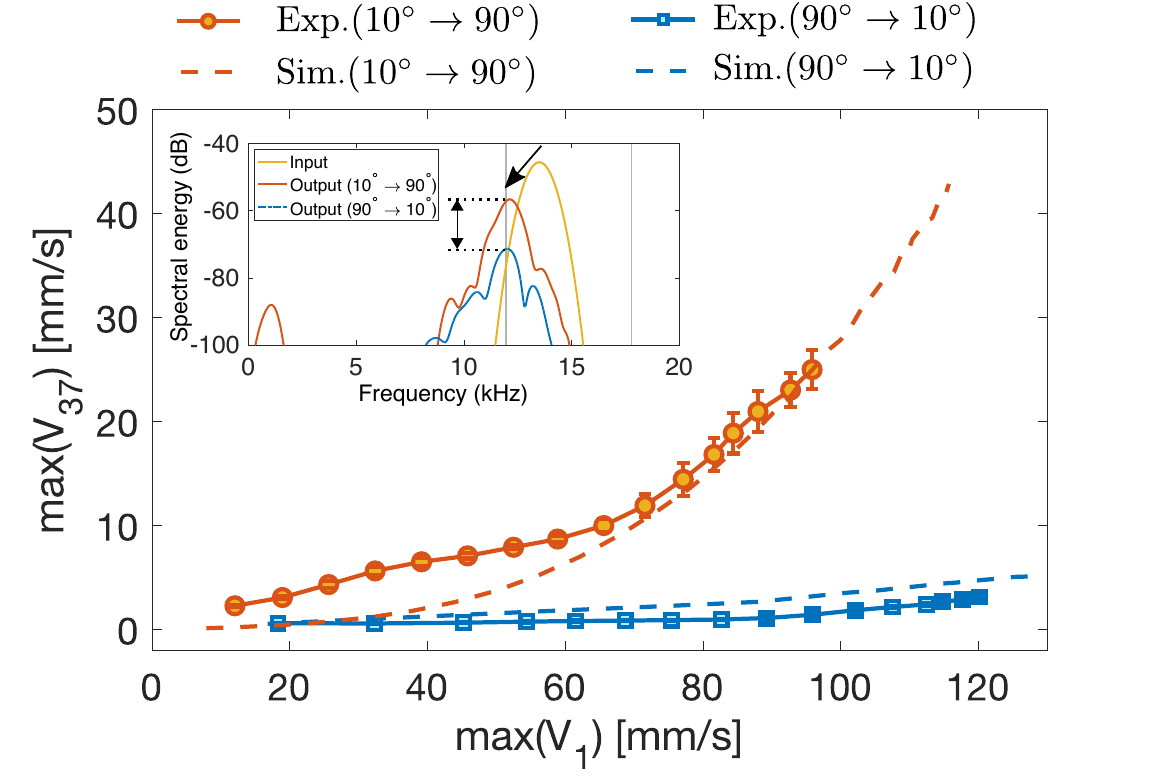}
\caption{(Color online) Comparison of asymmetric wave transmission data obtained from experiments and numerical simulations for two configurations: forward ($10^{\circ} \rightarrow 90^{\circ}$) and reverse ($90^{\circ} \rightarrow 10^{\circ}$). The maximum velocity of the last (37th) particle  from the actuator is compared with various excitation amplitude of the first particle from the actuator. In inset is the experimentally-obtained frequency content at the output of the chain. Two vertical grey lines denote the region of local cutoff frequencies in the chain. The arrow indicates the frequency shift from the input to the lowest local cutoff frequency. Double-sided arrow denotes the magnitude of asymmetry.}
\label{fig5}
\end{figure}

Next, we experimentally demonstrate the asymmetric wave transmission in our gradient-index chain.
We send a Gaussian-modulated pulse used in Fig.~\ref{fig4} from the actuator to the two different configurations: forward ($10^{\circ} \rightarrow 90^{\circ}$) and reverse ($90^{\circ} \rightarrow 10^{\circ}$), and measure wave transmission.
In Fig.~\ref{fig5}, we show the experimental evidence of asymmetric transmission in our system when the excitation amplitude is increased. The numerical simulation, which also includes the effect of viscous damping, follows the experimental data with a decent agreement. The inset highlights the aforementioned frequency shift governed by the local cutoff frequencies, and thereby leading to asymmetric wave transmission of about 15 dB. Note that the excitation range in the experiments is narrower than that in the simulations due to the limitation of our piezoelectric stack actuator. However, the asymmetric transmission is clearly verified within the range covered. 

\textit{Conclusion}.---We have proposed a highly tunable gradient-index system that is made of cylindrical granules. The contact interaction allows us to easily maintain a stiffness gradient along the chain. Due to the nonlinear Hertz contact law, the system is further tunable by the amplitude of wave excitation. For small amplitudes, the system follows linear dynamics and shows three distinctive frequency regions of wave transmission. These are a stop band, a pass band, and a quasi stop band that allows waves to penetrate only to a fraction of the system and then return back to the point of excitation. We experimentally demonstrate such a boomerang motion. For high amplitude excitations, we invoke nonlinear effects (gradual frequency softening) in the system, and demonstrate that the same system supports asymmetric wave transmission, leading to a rapid enhancement of transmission from one end to the other.
Therefore, this contact-based tunable system can inspire
novel class of systems to manipulate the flow of elastic energy for engineering applications, e.g., impact mitigation, vibration filtering, energy harvesting, and even mechanical logic gates. 
The underlying nonlinear mechanism of our system can also stimulate future studies in other domains such as plasmonics and photonics.

\begin{acknowledgments}
We thank Panayotis Kevrekidis (University of Massachusetts, Amherst) and Georgios Theocharis (CNRS) for valuable suggestions. E.K. acknowledges the support from the National Research Foundation of Korea, NRF-2017R1C1B5018136. J.Y. is grateful for the support of the National Science Foundation under
Grant No. CAREER-1553202.
\end{acknowledgments}


\bibliographystyle{apsrev4-1}

\bibliography{references.bib}

\end{document}


\widetext
\title{\large Supporting Information \\
Gradient-index granular crystals: From boomerang motion to \\ asymmetric transmission of waves}

\maketitle
\renewcommand{\thefigure}{S\arabic{figure}}

\section{Numerical Modeling}
\subsection{Discrete Element Model}
We use the discrete element method to simulate the wave propagation in graded chains. We model each cylinder as a point mass moving longitudinally and connected with nonlinear springs with neighboring cylinders. This is reasonable because the lowest resonant frequency of the cylindrical particle is much \textit{higher} than the frequencies of interest (i.e., frequencies of propagating waves along the chain). Thus we can assume that the particle moves as a rigid point mass. 
The contact force between $i$th and ($i+1$)th cylinders that make the contact angle of $\alpha_i$ is given by $F=\beta(\alpha_i)(\delta_{i} + u_i -u_{i+1})^{3/2}$ with $\beta(\alpha_i)$ taking the following form~\citep{Johnson1985, Khatri2012}
\begin{eqnarray}
\begin{aligned}
\label{eqn5}
\beta(\alpha) =\frac{2Y}{3(1-\nu^2)} \sqrt{\frac{R}{\sin{{\alpha}}}} \Bigg[\frac{2K(\e)}{\pi}\Bigg]^{-3/2} 
  \Bigg\{\frac{4}{\pi {\e}^2} \sqrt{\bigg[\bigg(\frac{r_1}{r_2}\bigg)^2 E(\e)-K(\e)\bigg][K(\e)-E(\e)]} \Bigg\}^{1/2}. \nonumber
\end{aligned}
\end{eqnarray}

\noindent Here $E$, $\nu$ and $R$ represent Young's modulus, Poisson's ratio, and the radius of each cylinder, respectively. 
The elliptical contact area between the cylinders has eccentricity $\e=\sqrt{1-(r_2/r_1)^2}$, where $r_1$ and $r_2$ are semi-major and semi-minor axes, respectively. $K(\e)$ and  $E(\e)$ are the complete elliptical integrals of the first and second kind, respectively. We further assume $r_2/r_1 \approx  [(1+\cos{\alpha})/(1-\cos{\alpha})]^{-2/3}$~\citep{Johnson1985}.

\subsection{Equations of motion}
For a chain of $N=37$ cylinders of mass $m$, we thus write equations of motion as 
\begin{subequations}
\begin{align}
{m} \frac{\mathrm{d}^2 u_1}{\mathrm{d} t^2} &= {\beta _a} \left[{{\delta _{a}} + {u_a - u_1}}\right]_+^{3/2}- {\beta _1}\left[ {{\delta _1} + {u_1} - {u_{2}}} \right]_ + ^{3/2} \nonumber -\frac{m}{\tau} \frac{\mathrm{d} u_1}{\mathrm{d} t}, \nonumber  \\
{m} \frac{\mathrm{d}^2 u_i}{\mathrm{d} t^2} &= {\beta _{i-1}}\left[ {{\delta _{i - 1}} + {u_{i - 1}} - {u_i}} \right]_ + ^{3/2} - {\beta _{i}}\left[ {{\delta _i} + {u_i} - {u_{i + 1}}} \right]_ +^{3/2}  \nonumber - \frac{m}{\tau} \frac{\mathrm{d} u_i}{\mathrm{d} t}, \nonumber \\
{m} \frac{\mathrm{d}^2 u_N}{\mathrm{d} t^2} &= {\beta _{N-1}}\left[ {{\delta _{N- 1}} + {u_{N - 1}} - {u_N}} \right]_ + ^{3/2} - F_0 \nonumber  
 - \frac{m}{\tau} \frac{\mathrm{d} u_N}{\mathrm{d} t}, \nonumber
\end{align}
\end{subequations}

\noindent where $u_i$ is dynamic displacement of $i$th cylinder, $\delta_i$ and $\beta_i$ are the static compression and contact stiffness coefficient between $i$th and ($i+1$)th cylinders, respectively. $u_a$ represents the actuator displacement, which is Gaussian modulated such that $u_a(t)=A \exp{-(t-\mu)/2 \sigma^2} \sin(\omega t)$ with $A$, $\sigma$, and $\mu$ being the peak amplitude, RMS width, and the time of maximum displacement, respectively. 
The bracket $[x]_+ = \max(0,x)$ is to make sure that we do not consider a tensile force for the contacts. $\beta_a$ is the contact stiffness coefficient between the actuator tip and the first particle in the chain, and it is taken to be 16.97 N/$\SI{}{\micro\metre}^{3/2}$ after considering the material properties and geometric curvatures at the contact. For the last particle in the chain, we assume that the pre-compressive force equivalent to the weight of 3 kg mass is directly applied. To consider the effect of viscous dissipation, we use an effective damping model and apply a force of $-(m/\tau) (\mathrm{d} u_i / \mathrm{d} t)$ on each particle. We choose $\tau=0.6$ ms to match wave decay in the experiments. 

\section{Wave amplification}
\begin{figure}[t]
\centering
\includegraphics[width=7in]{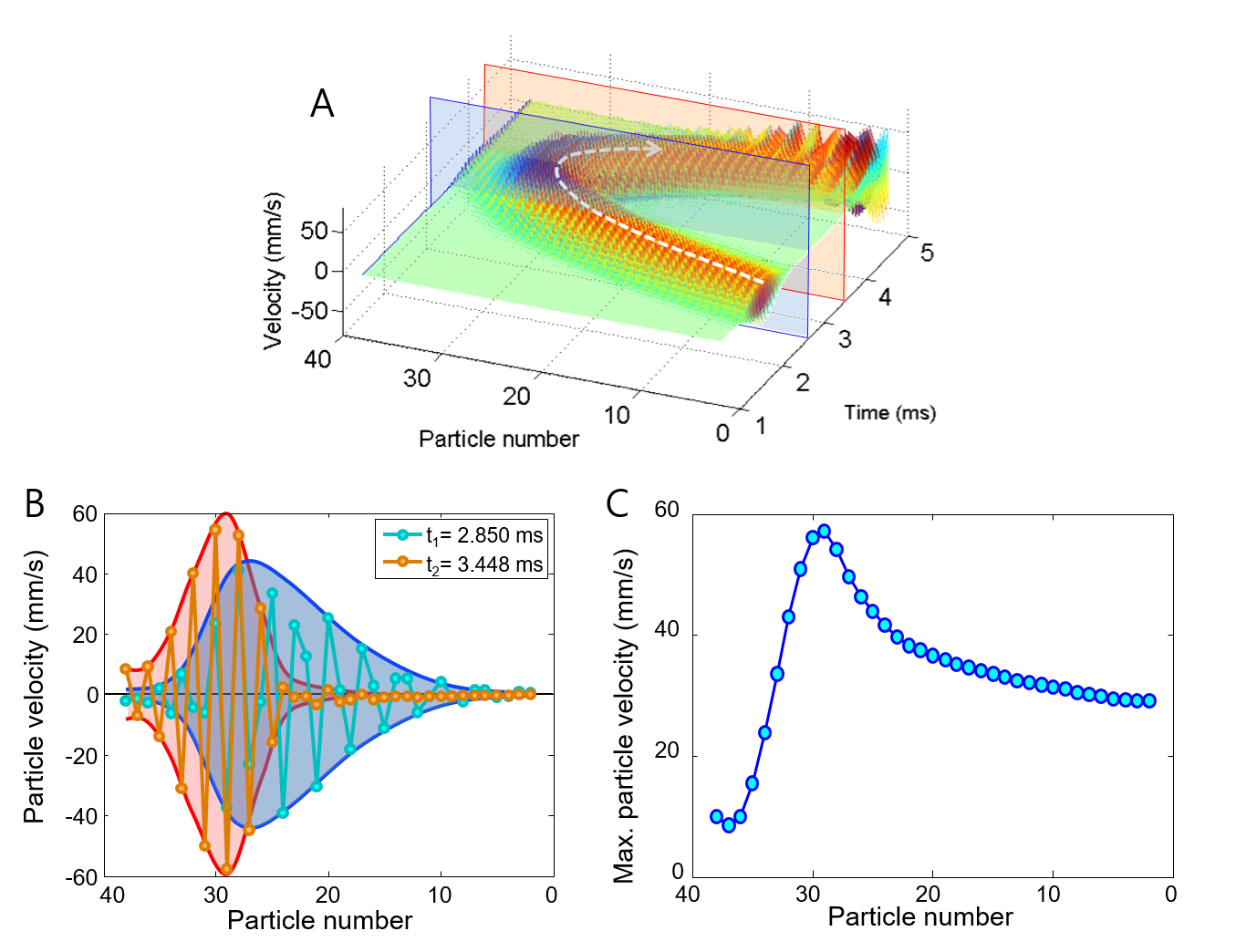}
\caption{(Color online) (A) Boomerang-like wave motion of a Gaussian wave packet with the center frequency of 13.5 kHz. The blue and red squares represent the moments at 2.850 ms and 3.448 ms, respectively. (B) Comparison of wave profiles at 2.850 ms and 3.448 ms. (C) Maximum particle velocity at each particle location.}
\label{figS1}
\end{figure} 

In this section, we discuss the surge of the wave amplitude in boomerang-like wave motion (see Fig.~\ref{figS1}A). A gradual variation in stiffness causes the wave speed to change along the length of the chain. For the configuration: $10^{\circ} \rightarrow 90^{\circ}$, the injected wave experiences decreasing stiffness, and therefore, the wave-packet's group velocity gradually decreases. 
The linear dynamics does not allow the change in the frequency, and therefore, the decrease in the group velocity directly implies an increase in the wavevector content. This manifests as the decrease in spatial width and an increase in amplitude of the wave packet as it traveling along the chain. We compare two wave profiles in Fig.~\ref{figS1}B. As the wave reaches the turning point, at about 3.448 ms, it is shrunk in width and amplified in amplitude. When the wave turns back, wave speed increases again and it is stretched in width. We show the maximum wave amplitude at each particle location in Fig.~\ref{figS1}C to further clarify this wave amplification effect.

\section{Asymmetric wave transmission: Frequency conversion}
\begin{figure}[t]
\centering
\includegraphics[width=7in]{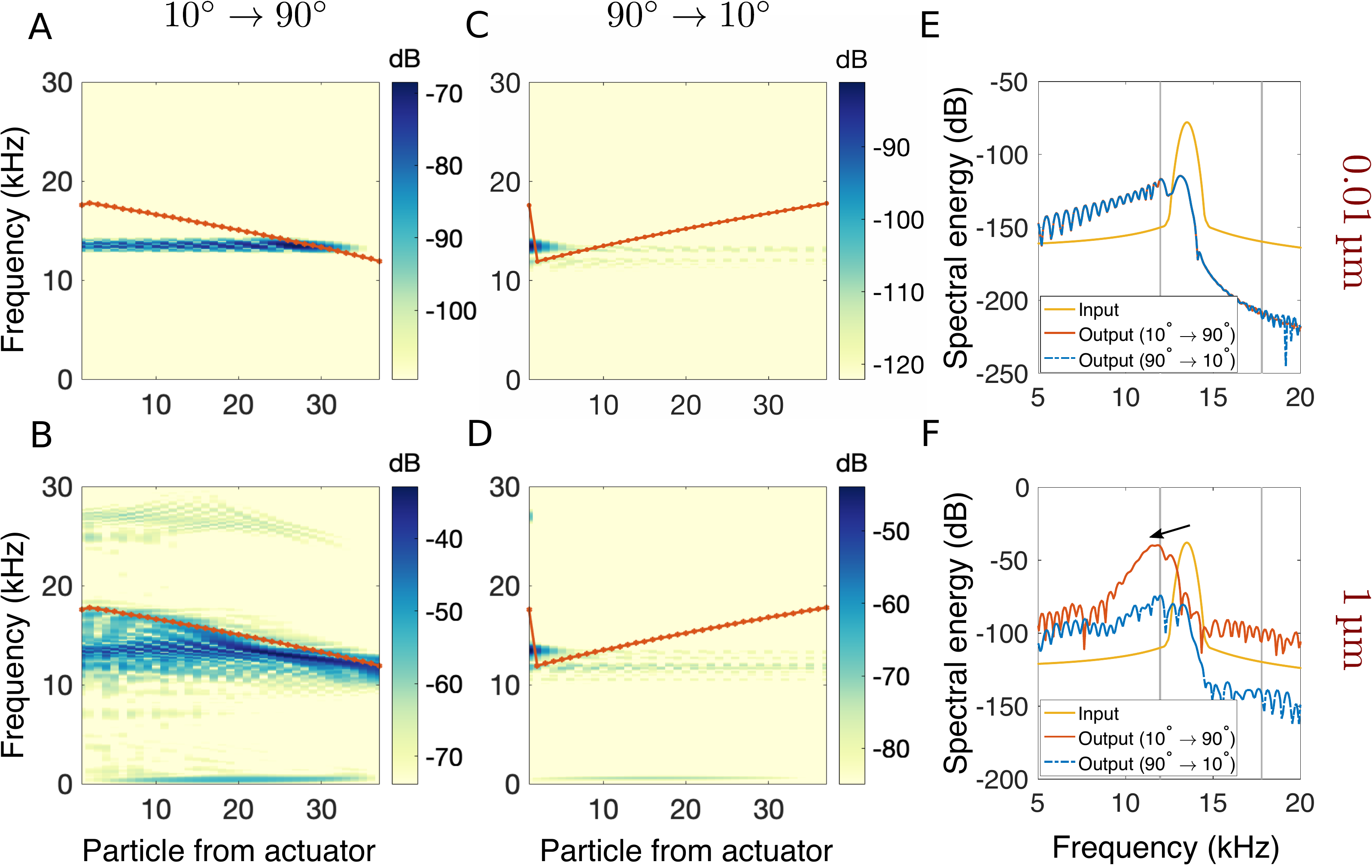}
\caption{(Color online) Wave spectrum of the chain excited at small and large amplitudes as described in Fig.~4 of the main text. (A)-(B) Frequency spectrum as a function of particle locations for forward ($10^{\circ} \rightarrow 90^{\circ}$) chain with small (\SI{0.01}{\micro\metre}) and large (\SI{1}{\micro\metre}) input excitation, respectively. The highlighted line in red represents analytically obtained local cutoff frequencies. (C)-(D) The same for the reverse ($90^{\circ} \rightarrow 10^{\circ}$) configuration. The step change in the local cutoff frequency near the actuator is due to the fact that the stiffness between the actuator and the first particle remains the same for both forward and reverse configurations. (E) Spectrum at the last particle for the configurations in (A) and (C), excited with small amplitude Gaussian pulse. Vertical grey lines denote the region of local cutoff frequencies in between. (F) Spectrum at the last particle for the configurations in (B) and (D), excited with large amplitude Gaussian pulse.  A frequency shift (black arrow) in the transmission, complying with the gradient, is apparent for the forward chain.}
\label{figS2}
\end{figure}

In this section, we discuss more on the frequency conversion due to nonlinearity in the graded chain. Figure ~\ref{figS2}A shows frequency spectrum, calculated from velocity time-history at each particle, for the forward chain ($10^{\circ} \rightarrow 90^{\circ}$). We use a Gaussian wave packet centered at 13.5 kHz as the input excitation, the same as in Fig.~4 of the main text and neglect viscous damping. The excitation amplitude is taken \SI{0.01}{\micro\metre} so that we invoke linear wave dynamics. We observe from the spectrum that the wave turns back from a point in the chain that is reasonably dictated by the local cutoff frequency plotted in red. When we further increase the excitation amplitude to \SI{1}{\micro\metre} in Fig.~\ref{figS2}B, nonlinear effects become apparent, and we observe that the spectral content of the wave gradually shifts downwards as the pulse travels away from the actuator, complying well with the gradual decrease in the local cutoff frequency. In addition, we observe some energy transfer to a low frequency domain (around 0 kHz) and a higher harmonic wave ($\approx$ 27 kHz).
We then perform a similar analysis on the reverse chain ($90^{\circ} \rightarrow 10^{\circ}$) and show the spectrum in Figs.~\ref{figS2}C and \ref{figS2}D for small and large amplitude excitation, respectively. In contrast to the forward chain, we do not observe any significant frequency conversion and transmission when excitation is increased. 

To quantify the transmission more clearly, in Fig.~\ref{figS2}E, we show the spectrum obtained at the output (the last particle) in comparison to the given small input excitation. We observe that the spectral energy at the output is strikingly similar for both forward and reverse configurations of the chain. Hence an asymmetric transmission is not evident as reported in Fig. 4 for small excitation amplitudes. However in Fig.~\ref{figS2}F, we clearly observe a downshift in frequency at the output of the forward chain. This shift is seen to conform with the local cutoff frequency at the output as also noted in Fig.~\ref{figS2}B. On the other hand, for the reverse chain, the spectral peak is much attenuated, and thus, it explains the asymmetric transmission reported in Fig. 4 for large amplitudes. We, therefore, conjecture that the asymmetric energy transmission is a result of exciting spatially-extended mode (gradon) in the forward configuration, which then shows \textit{gradual} frequency-softening (complying with local cutoff frequency) and becomes even more evident for large amplitude excitation. We leave the detailed mathematical treatment of this phenomena for future studies.

\section{The effect of stiffness gradient profile}

\begin{figure}[t]
\centering
\includegraphics[width=7in]{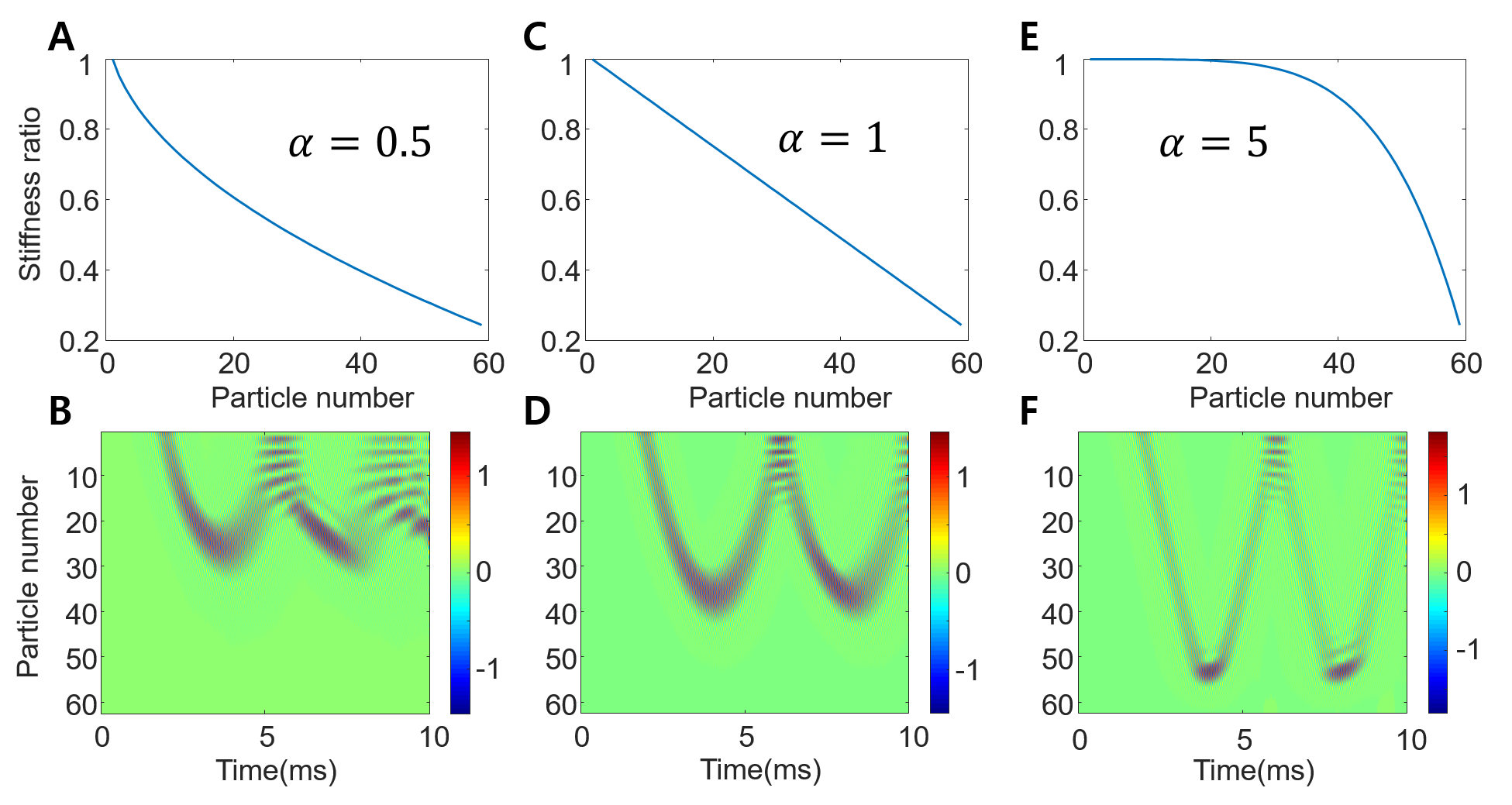}
\caption{(Color online) Boomerang behaviors in the cylinder chains with three different stiffness gradients following power laws.  A Gaussian wave centered at 15 kHz and with a maximum particle velocity 0.9 mm/s is sent from one end.  (A) Stiffness variation when the power factor $\alpha$ is 0.5 and (B) corresponding spatiotemporal evolution of wave packet. (C) Stiffness variation when $\alpha$ is 1 and (D) the corresponding spatiotemporal evolution. (E) Stiffness variation when $\alpha$ is 5 and (F) the corresponding spatiotemporal evolution.}
\label{figS3}
\end{figure}

In this section, we discuss the influence of stiffness gradient profile on the boomerang effect. Figure~\ref{figS3} shows boomerang motions in three systems with different stiffness configurations. They all following power laws, where the stiffness between the maximum ($k_{max}$) and the minimum ($k_{min}$) varies as below:

\begin{eqnarray}
\begin{aligned}
\label{eqn5}
k(i) =k_{max}-\frac{k_{max}-k_{min}}{(n-1)^{\alpha}}(i-1)^{\alpha},
\end{aligned}
\end{eqnarray}

where $i$ is the location index, $n$ is the total particle number,  and $ \alpha$ is the power law factor. In this way, the stiffness gradually decreases as the particle contact position (i.e., $i$) increases from $1$ to $n$. Figures~\ref{figS3} A,C, and E show the stiffness ratio ($k(i)/k_{max}$) along the chain length at three different powers ($\alpha = 0.5$, $\alpha = 1$, $\alpha = 5$). Figures~\ref{figS3} B,D, and F represent the corresponding wave motions obtained from numerical simulations. When the power is less than 1 (see Fig.S3 A \& B), the stiffness decreases fast in front part  of the chain. This makes the local cutoff frequencies of the system alter rapidly in the beginning of the chain, and thus the wave turns earlier compared to that of $\alpha = 1$ (see Figs. S3 C \& D). This is opposite when the power is larger than 1 (see Figs. S3 E \& D). In the case of $\alpha = 5$, we observe the waves propagate with a almost constant wave speed at the beginning, and then, it makes a rapid turn towards the end of the chain. The penetration depth is larger than the cases with lower $\alpha$ values. Overall, we observe that the stiffness gradient profile can be a fine parameter to control the boomerang motion and the asymmetric wave transmission in nonlinear regime.

\bibliographystyle{apsrev4-1}

\bibliography{references.bib}